\documentclass[preprintnumbers,amsmath,amssymbm,prd]{revtex4}
\usepackage{epsfig}
\usepackage{graphicx}

\begin{document}
\title{Charged Gauss-Bonnet black holes supporting non-minimally coupled scalar clouds: 
Analytic treatment in the near-critical regime}
\author{Shahar Hod}
\affiliation{The Ruppin Academic Center, Emeq Hefer 40250, Israel}
\affiliation{ }
\affiliation{The Hadassah Institute, Jerusalem 91010, Israel}
\date{\today}

\begin{abstract}
\ \ \ Recent numerical studies have revealed the physically intriguing fact that charged 
black holes whose charge-to-mass ratios are larger than 
the critical value 
$(Q/M)_{\text{crit}}=\sqrt{2(9+\sqrt{6})}/5$ can 
support hairy matter configurations which are made of scalar fields with a 
non-minimal negative coupling to the Gauss-Bonnet invariant of the curved spacetime. 
Using {\it analytical} techniques, we explore 
the physical and mathematical properties of the 
composed charged-black-hole-nonminimally-coupled-linearized-massless-scalar-field configurations 
in the near-critical $Q/M\gtrsim (Q/M)_{\text{crit}}$ regime. 
In particular, we derive an analytical resonance formula that describes the charge-dependence of 
the dimensionless coupling parameter $\bar\eta_{\text{crit}}=\bar\eta_{\text{crit}}(Q/M)$ 
of the composed Einstein-Maxwell-nonminimally-coupled-scalar-field system along the 
{\it existence-line} of the theory, a critical border that separates bald Reissner-Nordstr\"om black holes from 
hairy charged-black-hole-scalar-field configurations. 
In addition, it is explicitly shown that the large-coupling $-\bar\eta_{\text{crit}}(Q/M)\gg1$ analytical results derived in
the present paper for the composed Einstein-Maxwell-scalar theory agree remarkably well with direct numerical computations of the 
corresponding black-hole-field resonance spectrum.
\end{abstract}
\bigskip
\maketitle

\section{Introduction}

Early no-hair theorems \cite{Bek1,Sot2,Her1,Sot3,BekMay,Hod1}, 
which were mainly motivated by Wheeler's no-hair conjecture for black holes
\cite{NHC,JDB}, have revealed the physically interesting fact that 
black holes with spatially regular horizons cannot support static matter configurations which are made of 
scalar fields.  

However, recent studies \cite{Sot5,Sot1,GB1,GB2} (see also \cite{ChunHer,SotN,Hodsg1,Hodsg2}) 
have explicitly demonstrated that the no-hair conjecture 
can be violated in generalized Einstein-scalar field theories whose actions are characterized by 
a direct non-trivial coupling $f(\phi){\cal G}$ 
between the scalar field $\phi$ and the Gauss-Bonnet invariant ${\cal G}$ of the curved spacetime. 
In particular, it has been revealed in the physically important works \cite{Sot5,Sot1,GB1,GB2} that spatially regular 
matter configurations which are made of non-minimally coupled scalar fields can be supported in asymptotically flat 
black-hole spacetimes \cite{Notechar,Hersc1,Hersc2,Hodsc1,Hodsc2,Moh,Hodnrx,HodJP}. 

The physically intriguing phenomenon of spontaneous scalarization of black holes in generalized 
Einstein-Gauss-Bonnet-scalar field theories \cite{Sot5,Sot1,GB1,GB2,ChunHer,SotN,Hodsg1,Hodsg2} 
owe its existence to the presence of an effective spatially-dependent mass term 
of the linearized form $-\bar\eta{\cal G}$, 
which reflects the direct non-trivial coupling between the scalar field and the 
Gauss-Bonnet curvature invariant, in the Klein-Gordon wave equation of the supported scalar configurations 
[see Eq. (\ref{Eq10}) below]. The dimensionless physical parameter $\bar\eta$ of the composed field theory 
controls the strength of the direct non-minimal interaction between the scalar field and the spatially-dependent 
Gauss-Bonnet invariant of the curved spacetime. 

The spontaneous scalarization phenomenon of {\it charged} black holes in composed 
Einstein-Maxwell-Gauss-Bonnet-scalar-field theories has been explored, using numerical techniques, 
in the physically interesting works \cite{Brih,Hernn} (see \cite{Hodca,Done,Herrecnum,BeCo,ChunHer,SotN,Hodspg1,Hodspg2} for recent studies 
of the spontaneous scalarization phenomenon of asymptotically flat spinning black holes). 
Intriguingly, it has been revealed in \cite{Brih,Hernn} that the composed charged-black-hole-nonminimally-coupled-scalar-field system 
is characterized by a charge-dependent {\it existence-line} $\bar\eta=\bar\eta({\bar Q})$ \cite{NoteQb} which marks 
the onset of the spontaneous scalarization phenomenon in the generalized 
Einstein-Maxwell-scalar field theory. In particular, for a given value of the 
black-hole electric charge ${\bar Q}$, the critical existence line marks the boundary between 
bald Reissner-Nordstr\"om black holes 
and composed charged-black-hole-nonminimally-coupled-scalar-field hairy configurations. 

The charge-dependent critical existence-line $\bar\eta=\bar\eta({\bar Q})$ of the generalized 
Einstein-Maxwell-scalar theory is composed of charged Reissner-Nordstr\"om black holes 
that support scalar `clouds' \cite{Hodlit,Herlit}, spatially regular matter configurations 
which are made of the non-trivially coupled linearized scalar fields. The characteristic existence-line of the physical system 
is universal in the sense that different non-trivially coupled Einstein-Maxwell-scalar field theories that 
share the same weak-field functional behavior
$f(\phi)=1+{\bar\eta}\phi^2/2+O(\phi^4)$ \cite{Brih,Hernn} of the coupling function are 
characterized by the same critical boundary between bald and hairy black-hole spacetimes. 

Interestingly, the numerical results presented in \cite{Brih,Hernn} have revealed that the spontaneous scalarization phenomenon 
of black holes in composed Einstein-Maxwell-scalar field theories with {\it negative} values of the non-minimal 
coupling parameter $\bar\eta$ may be 
induced by the electric charge of the supporting black hole. 
In particular, one finds that, in the $\bar\eta<0$ regime, the onset of the spontaneous scalarization phenomenon is 
marked by the dimensionless black-hole electric charge \cite{Brih,Hernn}
\begin{equation}\label{Eq1}
{\bar Q}_{\text{crit}}={{\sqrt{2(9+\sqrt{6})}}\over{5}}\  .
\end{equation}
Only charged black holes with ${\bar Q}\geq {\bar Q}_{\text{c}}$ 
can support non-minimally coupled spatially regular scalar clouds in the negative coupling $\bar\eta<0$ regime. 
Intriguingly, it has been demonstrated numerically in \cite{Brih,Hernn} that, 
in the near-critical regime ${\bar Q}/{\bar Q}_{\text{crit}}\to 1^+$, the hairy charged black holes 
are characterized by the large-coupling asymptotic relation 
\begin{equation}\label{Eq2}
-\bar\eta\to \infty\ \ \ \ \text{for}\ \ \ \ {\bar Q}/{\bar Q}_{\text{crit}}\to 1^+\  .
\end{equation}

The main goal of the present paper is to study, using {\it analytical} techniques, 
the physical and mathematical properties of the 
composed charged-Reissner-Nordstr\"om-black-hole-nonminimally-coupled-scalar-field cloudy configurations 
in the near-critical regime ${\bar Q}\gtrsim {\bar Q}_{\text{crit}}$. 
In particular, a remarkably compact WKB resonance 
formula, which describes the charge-dependence $\bar\eta=\bar\eta({\bar Q})$ of 
the characteristic critical existence-line of the composed Einstein-Maxwell-Gauss-Bonnet-nonminimally-coupled-massless-scalar-field theory 
in the dimensionless large-coupling $-\bar\eta\gg1$ regime, will be derived. 

Interestingly, we shall explicitly show below that 
the analytically derived near-critical resonance formula of the present paper [see Eq. (\ref{Eq40}) below] 
provides a simple {\it analytical}
explanation for the {\it numerically} observed \cite{Brih,Hernn} monotonically decreasing 
functional behavior $|\bar\eta|=|\bar\eta({\bar Q})|$ of the critical 
existence-line that characterizes, in the $\bar\eta<0$ regime, the composed 
charged-Reissner-Nordstr\"om-black-hole-nonminimally-coupled-linearized-scalar-field configurations. 

\section{Description of the system}

We explore the physical and mathematical properties of spatially regular scalar `clouds' 
(linearized scalar field configurations) which are supported by 
charged Reissner-Nordstr\"om black holes.
The supported massless scalar fields are characterized by a direct non-trivial (non-minimal) coupling 
to the Gauss-Bonnet invariant of the curved charged spacetime. 
The action of the composed Einstein-Maxwell-Gauss-Bonnet-nonminimally-coupled-massless-scalar-field 
system is given by the expression \cite{Hernn,Noteun}
\begin{equation}\label{Eq3}
S=\int
d^4x\sqrt{-g}\Big[{1\over4}R-{1\over4}F_{\alpha\beta}F^{\alpha\beta}-{1\over2}\nabla_{\alpha}\phi\nabla^{\alpha}\phi+f(\phi){\cal
G}\Big]\  ,
\end{equation}
where 
\begin{equation}\label{Eq4}
{\cal G}\equiv R_{\mu\nu\rho\sigma}R^{\mu\nu\rho\sigma}-4R_{\mu\nu}R^{\mu\nu}+R^2\
\end{equation}
is the Gauss-Bonnet curvature invariant. 

Following \cite{GB1,GB2,Hersc1}, we assume that the scalar function $f(\phi)$, which controls 
the non-trivial direct coupling of the scalar field to the Gauss-Bonnet curvature invariant, 
is characterized by the leading order universal functional behavior
\begin{equation}\label{Eq5}
f(\phi)={1\over2}\eta\phi^2\
\end{equation}
in the weak-field regime. 
As discussed in \cite{GB1,GB2,Hersc1}, the functional behavior (\ref{Eq5}) of the scalar coupling function in the 
weak scalar field regime guarantees that the Einstein-matter field equations 
are satisfied by the familiar scalarless black-hole solutions of general relativity (in our case, the Reissner-Nordstr\"om black-hole 
spacetime) in the $\phi\to0$ limit.
The physical parameter $\eta$ \cite{Noteetaa}, 
which may take either positive or negative values \cite{Brih,Hernn}, determines 
the strength of the direct (non-minimal) interaction between the spatially regular massless scalar field configurations 
and the Gauss-Bonnet invariant (\ref{Eq4}) of the charged curved spacetime. 

The supporting charged Reissner-Nordstr\"om black-hole spacetime 
is characterized by the line element \cite{ThWe,Chan,Notebl}
\begin{eqnarray}\label{Eq6}
ds^2=-{{\Delta}\over{r^2}}dt^2+{{r^2}\over{\Delta}}dr^2+r^2d\theta^2+r^2\sin^2\theta d\phi^2\  ,
\end{eqnarray}
where the metric function $\Delta$ is given by the functional expression 
\begin{equation}\label{Eq7}
\Delta\equiv r^2-2Mr+Q^2\  .
\end{equation}
The physical parameters $\{M,Q\}$ are respectively the mass and electric charge 
of the central supporting Reissner-Nordstr\"om black hole. 
The horizon radii of the curved black-hole spacetime (\ref{Eq6}) are determined by the 
zeros of the metric function $\Delta$:
\begin{equation}\label{Eq8}
r_{\pm}=M\pm(M^2-Q^2)^{1/2}\  .
\end{equation}
The radially-dependent Gauss-Bonnet invariant, 
which characterizes the curved Reissner-Nordstr\"om black-hole spacetime (\ref{Eq6}), 
is given by the expression \cite{Hernn}
\begin{equation}\label{Eq9}
{\cal G}_{\text{RN}}(r;M,Q)={{8}\over{r^8}}\big(6M^2r^2-12MQ^2r+5Q^4\big)\  .
\end{equation}

A variation of the action (\ref{Eq3}) with respect to the scalar field yields 
the generalized Klein-Gordon equation \cite{Hernn}
\begin{equation}\label{Eq10}
\nabla^\nu\nabla_{\nu}\phi=\mu^2_{\text{eff}}\phi\
\end{equation}
for the non-minimally coupled scalar field with the radially-dependent effective mass term 
\begin{equation}\label{Eq11}
\mu^2_{\text{eff}}(r;M,Q)=-\eta{\cal G}\  .
\end{equation}
This effective mass term reflects the direct (non-minimal) 
coupling of the scalar field $\phi$ to the Gauss-Bonnet invariant ${\cal G}$ [see Eq. (\ref{Eq9})] 
of the charged curved spacetime.

Interestingly, one finds that, in the regime \cite{Hernn,NoteQrp} 
\begin{equation}\label{Eq12}
Q\geq Q_{\text{crit}}=M\cdot{{\sqrt{2(9+\sqrt{6})}}\over{5}}\  ,
\end{equation}
the charged-dependent 
Gauss-Bonnet curvature invariant becomes negative in the exterior region 
\begin{equation}\label{Eq13}
r_+\leq r\leq \Big(1+{{1}\over{\sqrt{6}}}\Big)\cdot {{Q^2}\over{M}}\ 
\end{equation}
of the charged black-hole spacetime. 
Thus, the spatially-dependent effective mass term (\ref{Eq11}) with $\eta<0$ may become negative 
in the interval (\ref{Eq13}). As we shall explicitly prove below, this intriguing property of the 
coupled Einstein-Maxwell-nonminimally-coupled-scalar-field system (\ref{Eq3}) 
allows the central charged Reissner-Nordstr\"om black hole (\ref{Eq6}) to support bound-state linearized cloudy configurations 
of the non-minimally coupled scalar field.

Using the functional field decomposition \cite{Notelmlm}
\begin{equation}\label{Eq14}
\phi(r,\theta,\varphi)=\sum_{lm} R_{lm}(r)Y_{lm}(\theta,\varphi)\  ,
\end{equation}
one obtains from Eq. (\ref{Eq10}) the differential equation \cite{Hernn,Noteomt,Notecomr}
\begin{eqnarray}\label{Eq15}
{{d}\over{dr}}\Big(\Delta{{dR}\over{dr}}\Big)-l(l+1)R+
\eta\Big({{48M^2}\over{r^4}}-{{96MQ^2}\over{r^5}}+{{40Q^4}\over{r^6}}\Big)R=0\
\end{eqnarray}
for the radial part of the linearized non-minimally coupled scalar field in the supporting curved black-hole spacetime (\ref{Eq6}). 

The ordinary differential equation (\ref{Eq15}) determines the spatial behavior of the 
static non-minimally coupled linearized massless scalar field configurations in the supporting 
charged Reissner-Nordstr\"om black-hole spacetime (\ref{Eq6}). 
Following \cite{Brih,Hernn}, we shall assume that the scalar eigenfunction $\psi(r)$ of the 
bound-state field configurations is spatially well behaved with 
the physically motivated boundary conditions \cite{Brih,Hernn}
\begin{equation}\label{Eq16}
\psi(r=r_{\text{H}})<\infty\ \ \ \ ; \ \ \ \ \psi(r\to\infty)\to0\
\end{equation}
at the black-hole horizon and at spatial infinity. 

In the next section we shall study, using analytical techniques, the physical and mathematical 
properties of the negatively coupled charged-black-hole-nonminimally-coupled-linearized-massless-scalar-field configurations that characterize the composed Einstein-Maxwell-scalar field theory (\ref{Eq3}). 
In particular, we shall explicitly prove that the composed Reissner-Nordstr\"om-black-hole-scalar-field cloudy configurations 
in the $Q>Q_{\text{crit}}$ regime [see Eq. (\ref{Eq12})] with $\eta<0$ 
owe their existence to the presence of an effective near-horizon binding potential well [see Eq. (\ref{Eq31}) below].

\section{Composed Reissner-Nordstr\"om-black-hole-nonminimally-coupled-scalar-field configurations: A WKB analysis}

In the present section we shall determine the charge-dependent functional behavior $\eta=\eta(Q/M)$ of the 
critical existence-line, which characterizes the composed charged-black-hole-nonminimally-coupled-scalar-field cloudy configurations, 
in the dimensionless large-coupling regime
\begin{equation}\label{Eq17}
-\bar\eta\equiv -{{\eta}\over{M^2}}\gg1\  .
\end{equation}

Defining the new radial eigenfunction
\begin{equation}\label{Eq18}
\psi\equiv rR\
\end{equation}
and using the coordinate $y(r)$, which is defined by the radial differential relation \cite{Noteppmm}
\begin{equation}\label{Eq19}
dy={{r^2}\over{\Delta}}\cdot dr\  ,
\end{equation}
one obtains from equation (\ref{Eq15}) the Schr\"odinger-like radial equation
\begin{equation}\label{Eq20}
{{d^2\psi}\over{dy^2}}-V\psi=0\  ,
\end{equation}
where the radially-dependent potential $V[r(y)]$ of the composed 
charged-Reissner-Nordstr\"om-black-hole-nonminimally-coupled-scalar-field system is given by the functional expression 
\begin{eqnarray}\label{Eq21}
V(r;M,Q,l,\bar\eta)=h(r)\Big[{{l(l+1)}\over{r^2}}+{{2M}\over{r^3}}-{{2Q^2}\over{r^4}}\Big]+V_{\text{GB}}(r)\
\end{eqnarray}
with [see Eq. (\ref{Eq7})]
\begin{equation}\label{Eq22}
h(r)\equiv {{\Delta}\over{r^2}}=1-{{2M}\over{r}}+{{Q^2}\over{r^2}}\  .
\end{equation}
The presence of the non-trivial term 
\begin{equation}\label{Eq23}
V_{\text{GB}}(r;M,Q)=-\bar\eta\cdot {{h(r)}\over{r^2}}\Big({{48M^4}\over{r^4}}-{{96M^3Q^2}\over{r^5}}+{{40M^2Q^4}\over{r^6}}\Big)\
\end{equation}
in the effective interaction potential (\ref{Eq21}) is a direct consequence of the non-trivial (non-minimal) 
coupling between the Gauss-Bonnet curvature invariant (\ref{Eq9}) of the charged spacetime (\ref{Eq6}) and 
the supported scalar field. 

We shall now prove that, in the dimensionless large-coupling regime (\ref{Eq17}), 
the ordinary differential equation (\ref{Eq20}), 
which characterizes the composed Einstein-Maxwell-scalar field theory (\ref{Eq3}), is
amenable to an analytical treatment. 
In particular, a standard second-order WKB analysis of the Schr\"odinger-like ordinary differential 
equation (\ref{Eq20}) yields the familiar quantization condition \cite{WKB1,WKB2,WKB3}
\begin{equation}\label{Eq24}
\int_{y_{t-}}^{y_{t+}}dy\sqrt{-V(y;M,Q,\bar\eta)}=\big(n+{1\over2}\big)\cdot\pi\
\ \ \ ; \ \ \ \ n=0,1,2,...\
\end{equation}
for the discrete spectrum $\{\bar\eta(M,Q;n\}^{n=\infty}_{n=0}$ of the dimensionless coupling parameter which 
characterizes the composed charged-Reissner-Nordstr\"om-black-hole-nonminimally-coupled-scalar-field cloudy configurations. 
Here $n\in\{0,1,2,...\}$ is the discrete resonant parameter of the physical system. 
The critical existence-line of the field theory is determined by the fundamental $n=0$ mode. 
The integration limits $\{y_{t-},y_{t+}\}$ in the WKB integral relation (\ref{Eq24}) are the classical turning points of the effective 
binding potential (\ref{Eq21}). 
Using Eq. (\ref{Eq19}), one finds that the WKB relation (\ref{Eq24}) can be expressed in the integral form
\begin{equation}\label{Eq25}
\int_{r_{t-}}^{r_{t+}}dr{{\sqrt{-V(r;M,Q,\bar\eta)}}\over{h(r)}}=\big(n+{1\over2}\big)\cdot\pi\
\ \ \ ; \ \ \ \ n=0,1,2,...\  .
\end{equation}

Defining the dimensionless physical parameters 
\begin{equation}\label{Eq26}
Q\equiv Q_{\text{crit}}\cdot(1+\epsilon)\ \ \ \ ; \ \ \ \ \epsilon\geq0
\end{equation}
and
\begin{equation}\label{Eq27}
r\equiv r_+\cdot(1+x)\ \ \ \ ; \ \ \ \ x\geq0\  ,
\end{equation}
one obtains the near-critical ($\epsilon\ll1$) near-horizon ($x\ll1$) relations 
[see Eqs. (\ref{Eq8}), (\ref{Eq12}), and (\ref{Eq22})] 
\begin{equation}\label{Eq28}
{{r_+}\over{M}}={{4+\sqrt{6}}\over{5}}\cdot\big(1-\sqrt{6}\cdot\epsilon\big)
+O(\epsilon^2)\  ,
\end{equation}
\begin{equation}\label{Eq29}
{{r}\over{M}}={{r_+\cdot(1+x)}\over{M}}={{4+\sqrt{6}}\over{5}}\cdot\big(1+x-\sqrt{6}\cdot\epsilon\big)
+O(x^2,\epsilon^2,x\epsilon)\  ,
\end{equation}
and
\begin{equation}\label{Eq30}
h(r)=(\sqrt{6}-2)\cdot x
+O(x^2,\epsilon^2,x\epsilon)\  .
\end{equation}

Substituting the relations (\ref{Eq28}), (\ref{Eq29}), and (\ref{Eq30}) into Eqs. (\ref{Eq21}) and (\ref{Eq23}), 
one finds the (rather cumbersome) near-critical near-horizon expression
\begin{equation}\label{Eq31}
M^2{{V(r)}\over{h(r)}}=\Bigg\{{{11-4\sqrt{6}}\over{2}}\cdot l(l+1)+{{19\sqrt{6}-46}\over{2}}-
\bar\eta\Big[\big(15204\sqrt{6}-37236\big)\cdot x-\big(16752-6828\sqrt{6}\big)\cdot\epsilon\Big]\Bigg\}\cdot[1+O(x,\epsilon)]\  
\end{equation}
for the effective interaction potential of the composed black-hole-massless-scalar-field cloudy configurations.

We shall henceforth consider composed 
charged-Reissner-Nordstr\"om-black-hole-massless-scalar-field cloudy configurations 
in the dimensionless large-coupling regime [see Eqs. (\ref{Eq17}), (\ref{Eq26}), and Eq. (\ref{Eq38}) below]
\begin{equation}\label{Eq32}
-\bar\eta\epsilon\gg1\  ,
\end{equation}
in which case one finds from Eqs. (\ref{Eq30}) and (\ref{Eq31}) the remarkably compact leading order functional expression 
\begin{equation}\label{Eq33}
M^2{{V(r)}\over{[h(r)]^2}}={\bar\eta}
\cdot6(1396-569\sqrt{6})\cdot\Big[(2+\sqrt{6})\cdot{{\epsilon}\over{x}}-1\Big]\
\end{equation}
for the effective binding potential of the composed charged-black-hole-scalar-field system. 

Taking cognizance of Eqs. (\ref{Eq25}), (\ref{Eq27}), and (\ref{Eq33}), one finds the resonance condition
\begin{equation}\label{Eq34}
{{r_+}\over{M}}\cdot\int_{0}^{(2+\sqrt{6})\cdot\epsilon}dx \sqrt{-{\bar\eta}
\cdot6(1396-569\sqrt{6})\cdot\Big[(2+\sqrt{6})\cdot{{\epsilon}\over{x}}-1\Big]
}=\big(n+{1\over2})\cdot\pi\ \ \ \ ; \ \ \
\ n=0,1,2,...\  
\end{equation}
for the composed black-hole-field system. 
The WKB integral relation (\ref{Eq34}) can be expressed in the compact mathematical form
\begin{equation}\label{Eq35}
\epsilon\sqrt{-\bar\eta}\cdot2\sqrt{6}\sqrt{16+\sqrt{6}}\int_{0}^{1}dz
\sqrt{{{1}\over{z}}-1}=\big(n+{1\over2})\cdot\pi\ \ \ \ ; \ \ \ \
n=0,1,2,...\  ,
\end{equation}
where
\begin{equation}\label{Eq36}
z\equiv {{1}\over{2+\sqrt{6}}}\cdot{{x}\over{\epsilon}}\  .
\end{equation}

Using the integral relation 
\begin{equation}\label{Eq37}
\int_{0}^{1}dz \sqrt{{{1}\over{z}}-1}={{\pi}\over{2}}\  ,
\end{equation}
one finds from (\ref{Eq35}) the remarkably simple discrete resonance formula \cite{Notegdp}
\begin{equation}\label{Eq38}
\bar\eta(\epsilon;n)=-{{1}\over{\epsilon^{2}}}\cdot{{1}\over{96+6\sqrt{6}}}
\cdot\big(n+{1\over2})^2\ \ \ \ ; \ \ \ \ n=0,1,2,...\  ,
\end{equation}
which characterizes the composed 
charged-Reissner-Nordstr\"om-black-hole-nonminimally-coupled-linearized-massless-scalar-field cloudy configurations in 
the dimensionless large-coupling $-\bar\eta\gg1$ regime (or equivalently, in the near-critical $\epsilon\ll1$ regime).

The discrete resonance spectrum (\ref{Eq38}) of the composed 
Einstein-Maxwell-Gauss-Bonnet-nonminimally-coupled-scalar-field system (\ref{Eq3}) 
can be expressed in the dimensionless form [see Eqs. (\ref{Eq1}) and (\ref{Eq26})]
\begin{equation}\label{Eq39}
{\bar Q}(\bar\eta;n)={{\sqrt{2(9+\sqrt{6})}}\over{5}}+
{{1}\over{\sqrt{-\bar\eta}}}\cdot{{1}\over{\sqrt{138-7\sqrt{6}}}}\cdot\big(n+{1\over2})\ \ \ \ ; \ \ \ \
n=0,1,2,...\  ,
\end{equation}
where ${\bar Q}\equiv Q/M$. 

\section{Numerical confirmation}

In the present section we shall test the accuracy of the analytically
derived large-coupling resonance spectrum (\ref{Eq38}), which characterizes the composed 
charged-Reissner-Nordstr\"om-black-hole-nonminimally-coupled-linearized-scalar-field 
cloudy configurations. The charge-dependent resonance spectrum of the black-hole-field system 
has recently been computed numerically in \cite{Hernn}. 

In Table \ref{Table1} we present, for various values of the dimensionless coupling parameter 
$\lambda\equiv2\sqrt{{|\bar\eta|}}$ \cite{Notecomr} used in \cite{Hernn}, the charge-dependent 
ratio ${\cal R}(\lambda)\equiv {\epsilon}^{\text{analytical}}/{\epsilon}^{\text{numerical}}$ 
between the analytically calculated value of the dimensionless critical parameter $\epsilon$ [as calculated directly 
from the analytically derived large-coupling resonance formula (\ref{Eq38})] and the 
corresponding exact (numerically computed \cite{Hernn}) values of the critical parameter. 
The data presented in Table \ref{Table1} for the cloudy Reissner-Nordstr\"om-black-hole-scalar-field configurations 
reveals the fact that the agreement between the {\it analytically} derived resonance formula (\ref{Eq38}) 
and the corresponding {\it numerically} computed resonance spectrum of \cite{Hernn} is remarkably good 
in the large-coupling $\lambda\gg1$ regime 
\cite{Noteetaeps} of the Einstein-Maxwell-scalar field theory (\ref{Eq3}). 

\begin{table}[htbp]
\centering
\begin{tabular}{|c|c|c|c|c|c|}
\hline \ \ $\lambda$\ \ & \ $5$\ \ & \ $10$\ \ & \
$15$\ \ & \ $20$\ \ & \ $\ 25$\ \ \\
\hline \ \ ${\cal R}(\lambda)\equiv {{{\epsilon}^{\text{analytical}}}\over{{\epsilon}^{\text{numerical}}}}$\ \ &\ \
$1.168$\ \ \ &\ \ $1.065$\ \ \ &\ \ $1.049$\ \ \ &\ \ $1.019$\ \ \ &\ \ $1.016$\ \ \\
\hline
\end{tabular}
\caption{Composed charged-Reissner-Nordstr\"om-black-hole-nonminimally-coupled-linearized-massless-scalar-field 
cloudy configurations. We present, for various values of the coupling parameter 
$\lambda\equiv2\sqrt{{|\bar\eta|}}$ \cite{Hernn,Notecomr} of the theory, 
the dimensionless ratio ${\cal R}(\lambda)\equiv {\epsilon}^{\text{analytical}}/{\epsilon}^{\text{numerical}}$ 
between the analytically calculated value of the dimensionless critical parameter $\epsilon$ [as calculated directly 
from the resonance formula (\ref{Eq38}) for the fundamental $n=0$ mode] 
and the corresponding exact values of the critical parameter as computed numerically in \cite{Hernn}. 
One finds that, in the large-coupling $\lambda\gg1$ regime of 
the composed Reissner-Nordstr\"om-black-hole-scalar-field cloudy configurations, 
the agreement between the analytically derived resonance formula (\ref{Eq38}) 
and the corresponding numerically computed resonance spectrum of \cite{Hernn} is remarkably good \cite{Noteetaeps}.} \label{Table1}
\end{table}

\section{Summary and Discussion}

Asymptotically flat black holes with spatially regular horizons can support bound-state matter configurations which 
are made of scalar fields with a direct (non-minimal) coupling to the Gauss-Bonnet invariant of the curved 
spacetime \cite{Sot5,Sot1,GB1,GB2,ChunHer,SotN,Hodsg1,Hodsg2}. 

The spontaneous scalarization phenomenon of charged black holes in composed 
Einstein-Maxwell-Gauss-Bonnet-scalar field theories has recently been studied numerically in the 
physically important works \cite{Brih,Hernn}. In particular, it has been revealed in \cite{Brih,Hernn} that 
a charge-dependent critical existence-line $\bar\eta_{\text{crit}}=\bar\eta_{\text{crit}}(Q/M)$ separates 
bald Reissner-Nordstr\"om black-hole spacetimes from the 
composed charged-black-hole-nonminimally-coupled-massless-scalar-field hairy configurations 
of the Einstein-Maxwell-scalar theory (\ref{Eq3}) \cite{Noteecr}, 
where the dimensionless physical parameter $\bar\eta$ quantifies the strength of the direct non-trivial interaction 
between the supported scalar field and the Gauss-Bonnet curvature invariant. 

Interestingly, it has been demonstrated \cite{Brih,Hernn} that, in the negative coupling $\bar\eta<0$ regime, 
the composed charged-black-hole-linearized-scalar-field cloudy configurations that sit on the 
critical existence-line of the Einstein-Maxwell-scalar field theory (\ref{Eq3}) are restricted to the dimensionless charge 
regime ${\bar Q}>{\bar Q}_{\text{crit}}={{\sqrt{2(9+\sqrt{6})}}/{5}}$ [see Eq. (\ref{Eq1})]. 
In particular, the numerical results presented in \cite{Brih,Hernn} provide important evidence for an intriguing 
divergent functional behavior $-\bar\eta({\bar Q})\to\infty$ 
of the non-minimal coupling parameter of the theory along the existence-line of the system 
in the near-critical limit ${\bar Q}/{\bar Q}_{\text{crit}}\to1^+$. 

In the present paper we have used {\it analytical} techniques in order to explore the physical and mathematical properties of the
composed charged-Reissner-Nordstr\"om-black-hole-nonminimally-coupled-massless-scalar-field hairy configurations of the Einstein-Maxwell-scalar theory (\ref{Eq3}) in the large-coupling $-\bar\eta({\bar Q})\gg1$ regime. 
In particular, using a WKB procedure we have derived the analytical formula (\ref{Eq38}) for the discrete 
resonant spectrum of the dimensionless coupling parameter $\bar\eta({\bar Q})$ which
characterizes the composed charged-black-hole-scalar-field cloudy
configurations in the near-critical $Q\gtrsim Q_{\text{crit}}$ regime. 

The analytically derived discrete WKB resonance formula (\ref{Eq38})
yields the remarkably compact charge-dependent functional relation \cite{Note00} 
\begin{equation}\label{Eq40}
\bar\eta(\epsilon)=-{{1}\over{\epsilon^{2}}}\cdot{{1}\over{4(96+6\sqrt{6})}}
\end{equation}
for the critical existence-line of the composed 
Einstein-Maxwell-Gauss-Bonnet-nonminimally-coupled-massless-scalar field theory (\ref{Eq3}), 
where $0\leq\epsilon=({{{\bar Q}-{\bar Q}_{\text{crit}}})/{{\bar Q}_{\text{crit}}}}\ll1$ [see Eqs. (\ref{Eq1}) and (\ref{Eq26})] 
is the dimensionless distance of the system from the exact critical ($-\bar\eta\to\infty$) configuration. 

It is worth stressing the fact that the physical significance of the critical existence-line (\ref{Eq40}) in 
the Einstein-Maxwell-scalar field theory (\ref{Eq3}) stems from the fact that it separates, 
in the large-coupling $-\bar\eta\gg1$ regime, bald Reissner-Nordstr\"om black holes from hairy 
charged-black-hole-nonminimally-coupled-massless-scalar-field configurations. From the analytically 
derived resonance formula (\ref{Eq40}) one finds, in accord with the physically interesting numerical results 
presented in \cite{Brih,Hernn}, that the dimensionless coupling parameter $|\bar\eta({\bar Q})|$ of the theory (\ref{Eq3}) 
is a monotonically decreasing function 
of the black-hole electric charge ${\bar Q}$ in the negative coupling $\bar\eta<0$ regime.

Finally, it is interesting to point out that our results provide a simple {\it analytical}
explanation for the physically intriguing {\it numerical} observation originally made in \cite{Brih,Hernn}, according to which 
the charge-dependent coupling parameter $\bar\eta({\bar Q})$ of the 
Einstein-Maxwell-scalar field theory (\ref{Eq3}) diverges along the critical existence line of the system 
in the $\bar Q\to {\bar Q}_{\text{crit}}$ limit. 
In particular, the analytically derived resonance formula (\ref{Eq40}) reveals the fact 
that the coupling parameter $\bar\eta(\bar Q)$, which characterizes the cloudy black-hole-field configurations, 
diverges quadratically in the near-critical $\epsilon\to0$ limit.

\bigskip
\noindent
{\bf ACKNOWLEDGMENTS}
\bigskip

This research is supported by the Carmel Science Foundation. I would
like to thank Yael Oren, Arbel M. Ongo, Ayelet B. Lata, and Alona B.
Tea for helpful discussions.

%\newpage


\begin{thebibliography}{99}
\bibitem{Notechar} A physically interesting related phenomenon, in which matter configurations that are made of 
scalar fields with a direct non-trivial (non-minimal) coupling 
to the Maxwell electromagnetic invariant are supported in 
charged black-hole spacetimes, has been revealed in \cite{Hersc1,Hersc2} (see also 
\cite{Hodsc1,Hodsc2,Moh,Hodnrx} and references therein).
\bibitem{NoteQb} Here ${\bar Q}\equiv Q/M$, where $\{M,Q\}$ are respectively the ADM mass and electric 
charge of the black-hole spacetime. We shall assume, without loss of generality, the relation ${\bar Q}>0$ for the black-hole electric charge.
\bibitem{Noteun} We use natural units in which $8\pi G=c=\hbar=1$.
\bibitem{Noteetaa} The coupling parameter $\eta$ of the composed field theory (\ref{Eq3}) has the dimensions of length$^2$.
\bibitem{Notebl} Here $(t,r,\theta,\phi)$ are the familiar Schwarzschild spacetime coordinates.
\bibitem{NoteQrp} Note that the critical electric charge (\ref{Eq12}) corresponds to the critical radius 
$r^c_+=M\cdot[(4+\sqrt{6})/5]$ of the black-hole outer horizon [see Eq. (\ref{Eq8})].
\bibitem{Notelmlm} Here $Y_{lm}(\theta,\varphi)$ are the spherical harmonic functions whose angular eigenvalues 
are given by the familiar expression $l(l+1)$ with the property $l\geq|m|$.
\bibitem{Noteomt} For brevity, we shall henceforth omit the angular field parameters $\{l,m\}$.
\bibitem{Notecomr} Note that the non-minimal coupling parameter $\eta$ of the composed 
Einstein-Maxwell-Gauss-Bonnet-nonminimally-coupled-massless-scalar-field theory (\ref{Eq3}) is 
related to the non-minimal coupling parameter $\lambda$ of \cite{Hernn} by the 
relation $\eta=\lambda^2/4$ [see Eq. (\ref{Eq15}) and Eq. (3.27) of \cite{Hernn}].
\bibitem{Noteppmm} Note that the differential relation (\ref{Eq19}) maps the 
semi-infinite regime $[r_+,\infty]$ of the Schwarzschild radial coordinate $r$ into the infinite 
regime $[-\infty,\infty]$ of the new radial coordinate $y$.
\bibitem{Notegdp} It is important to point out that, 
in accord with the assumed strong inequality (\ref{Eq32}), one finds from Eq. (\ref{Eq38}) the characteristic scaling relation 
$\bar\eta\epsilon=O(\epsilon^{-1})\gg1$ for the dimensionless coupling parameter of the 
composed Einstein-Maxwell-Gauss-Bonnet-nonminimally-coupled-massless-scalar-field theory 
in the near-critical $\epsilon\ll1$ regime. 
In particular, from the analytically derived resonance formula (\ref{Eq38}) one finds 
that, for the fundamental ($n=0$) resonant mode, the assumed strong inequality (\ref{Eq32}) is satisfied in the near-critical regime 
$\epsilon\ll[4(96+6\sqrt{6})]^{-1}$, or equivalently in the dimensionless large-coupling regime 
$-\bar\eta\gg4(96+6\sqrt{6})$.
\bibitem{Noteetaeps} Interestingly, from the data presented in Table \ref{Table1} one finds that 
the agreement between the analytically derived resonance formula (\ref{Eq38}) and the interesting 
numerical results presented in \cite{Hernn} for the cloudy black-hole-field configurations 
is quite good already in the $\bar\eta\epsilon=O(1)$ regime.
\bibitem{Noteecr} The critical existence-line, which characterizes the composed 
Einstein-Maxwell-Gauss-Bonnet-scalar field theory (\ref{Eq3}), corresponds to 
spatially regular non-minimally coupled linearized scalar field configurations (scalar clouds) that 
are supported by central charged Reissner-Nordstr\"om black holes.
\end{thebibliography}
\end{document}